\documentclass[aps, superscriptaddress, preprint]{revtex4-2}
\usepackage{amsmath}
\usepackage{amsfonts}
\usepackage{amssymb}
\usepackage{graphicx}
\usepackage{units}
\usepackage{color}
\usepackage{xspace}
\usepackage[normalem]{ulem}
\usepackage{natbib}
\usepackage{textcomp}

\usepackage{romannum}

\DeclareUnicodeCharacter{2009}{\,}
\DeclareUnicodeCharacter{2212}{-}

\begin{document}

\title{{Ultrahigh breakdown current density of van der Waals One Dimensional PdBr$_{\mathrm{2}}$}}

\author{Bikash Das}
\affiliation{School of Physical Sciences, Indian Association for the Cultivation of Science, 2A \& B
Raja S. C. Mullick Road, Jadavpur, Kolkata - 700032, India}

\author{Kapildeb Dolui}
\affiliation{Lomare Technolgies Limited, 6 London Street, London EC3R 7LP, United Kingdom}

\author{Rahul Paramanik}
\affiliation{School of Physical Sciences, Indian Association for the Cultivation of Science, 2A \& B
Raja S. C. Mullick Road, Jadavpur, Kolkata - 700032, India}

\author{Tanima Kundu}
\affiliation{School of Physical Sciences, Indian Association for the Cultivation of Science, 2A \& B
Raja S. C. Mullick Road, Jadavpur, Kolkata - 700032, India}

\author{Sujan Maity}
\affiliation{School of Physical Sciences, Indian Association for the Cultivation of Science, 2A \& B
Raja S. C. Mullick Road, Jadavpur, Kolkata - 700032, India}

\author{Anudeepa Ghosh}
\affiliation{School of Physical Sciences, Indian Association for the Cultivation of Science, 2A \& B
Raja S. C. Mullick Road, Jadavpur, Kolkata - 700032, India}

\author{Mainak Palit}
\affiliation{School of Physical Sciences, Indian Association for the Cultivation of Science, 2A \& B
Raja S. C. Mullick Road, Jadavpur, Kolkata - 700032, India}

\author{Subhadeep Datta}
\affiliation{School of Physical Sciences, Indian Association for the Cultivation of Science, 2A \& B
Raja S. C. Mullick Road, Jadavpur, Kolkata - 700032, India}

\email{sspsdd@iacs.res.in}



\begin{abstract}
One-dimensional (1D) van der Waals (vdW) materials  offer \textit{nearly defect-free} strands as channel material in the field-effect transistor (FET) devices and probably, a better interconnect than conventional copper with higher current density and resistance to electro-migration with sustainable down-scaling. We report a new halide based ``truly" 1D few-chain atomic thread, PdBr$_2$, isolable from its bulk which crystallizes in a monoclinic space group C2/c. Liquid phase exfoliated nanowires with mean length (20$\pm$1)$\mu$m  transferred onto SiO$_2$/Si wafer with a maximum aspect ratio (length:width) of $\approx 5000$ confirms the lower cleavage energy perpendicular to chain direction. Moreover, an isolated nanowire can also sustain current density of 200 MA/cm$^\mathrm{2}$ which is atleast one-order higher than typical copper interconnects. However, local transport measurement \textit{via} conducting atomic force microscopy (CAFM) tip along the cross direction of the single chain records a much lower current density due to the anisotropic electronic band structure. While 1D nature of the nanoobject can be linked with non-trivial collective quantum behavior, vdW nature could be beneficial for the new pathways in interconnect fabrication strategy with better control of placement in an integrated circuit (IC).
\end{abstract}

\maketitle

\section{Introduction}

In the vdW family, graphene \cite{Graphene} and graphene-like atomically thin two-dimensional (2D) layered materials (MoS$_2$, WS$_2$, MoTe$_2$, WTe$_2$, MoSe$_2$, WSe$_2$ \textit{etc.}), compared to its bulk equivalent, show distinct electrical, optical and magnetic properties  with unfolding the physics of semimetalicity, superconductivity, charge density wave, metal-insulator transition, \textit{etc.}\cite{MoS_2,WS_2,MoTe_2,WTe_2,MoSe_2}. Moreover, varying the carrier transport with electrostatic gating in ``all-2D" device brings forth electric field control over phase transition. However, one dimensional analog of these 2D vdW elements/compounds, barring carbon nanotube, did not achieve that much success due to complicated layer-by-layer ``\textit{in-situ}" growth template, imperfect phase purity and poor crystallinity. But, several non-vdW materials such as InAS, GaAs, InP, ZnO, $TiO_2$ etc., in their 1D forms, has been translated to prototypes like quantum dot laser \cite{QD_Science}, single-electron transistor \cite{Single Electron Transistor} , memory circuits \cite{Memory}, etc. 
Again, poor success rate of device fabrication with random performance distribution and lack of control over location-specific sticking on the desired substrate, hinders the practical routes towards volume manufacturing of non-vdW 1D materials. In such a case, easily exfoliable atomic chain like motif directly from the bulk crystal and facile transfer on the substrate for integration in nanoelectronic devices makes 1D vdW systems elusive. 

Recently, transition metal trichalcongenides (TMT) such as $TaSe_3$, $NbSe_3$, $TaS_3$, $TiS_3$, $ZrTe_3$, \textit{etc.} \cite{TaSe3, TiS3, Review1, Review2} have shown anisotropic electrical conductivity not only due to its quasi 1D vdW geometry but orbital composition of the bands. 
Besides, strain-induced removal of band inversion turns metallic $TaSe_3$, in its whisker crystal form, into topological superconductor \cite{TaSe3SC}. Also, halide based linear-chain compound, $(TaSe_4)_2I$, shows complex ground state with coexistence of superconductivity and ferromagnetism \cite{arnab}. Furthermore, as potential application, atomic scale antennas like $TaSe_3$ bundles in polymer composite  has been used as an efficient electromagnetic interference (EMI) shield \cite{TaSe3}. 
In device applications, Island \textit {et al.} reported a quasi-1D vdW \textit {n-type} FET with $TiS_3$ as transport channel having low electron mobility ($\sim$ 70 cm$^\mathrm{2}$V$^\mathrm{-1}$s$^\mathrm{-1}$) due to the influence of polar-phonon scattering \cite{TiS3, TiS3_2}. On a related note, nanowire of $ZrTe_3$, another material from vdW TMT family, shows high current carrying capacity ($\approx$ 100 MAcm$^{-2}$) for potential application in interconnect technologies \cite{ZrTe3}. Interestingly,  
Zhu \textit{et al.} identified a series of unexplored 1D vdW materials within the limits of exfoliability of typical 2D materials and predicted their electronic properties using density-functional-theory-based (DFT) methods \cite{Reed}. 

Here, we report one such predicted ``truly-1D" vdW material, palladium bromide ($PdBr_2$), which shows maximum aspect ratio of 5000 in its nanowire form, exfoliated from the bulk single crystals. The high value of lateral breakdown current density (200 MAcm$^{-2}$) indicates the potential use of the crystalline atomic chains as better interconnect than copper in microprocessors. Moreover, anisotropic electron transport in 1D channel based on current density in two mutually perpendicular direction of nanowire, measured \textit{via} conducting atomic force microscopy (CAFM), is corroborated with the band structure calculation.

\section{Experimental Details}

Bulk single crystals of PdBr$_2$ were grown by chemical vapor transport (CVT) method using PdBr$_2$ micropowder as a source (purity $99.99\%$, seller: Alfa Aesar). $200$ mg $PdBr_2$ powder was taken in a quartz tube which was evacuated down to $~10^{-5}$ torr and sealed under vacuum. The tube was placed in a horizontal two-zone tube furnace for 5 days, and the source and growth zone were increased to $700^{\circ} C$ and $400^{\circ} C$, respectively. After growth, the ampoule was cracked in an ambient atmosphere. Brown, shiny $PdBr_2$ crystals were collected from the middle and cold region of the tube, whereas polycrystalline $PdBr_2$ chunk was obtained at the hot zone of the tube.


Single crystal X-ray diffraction (SXRD) pattern was collected using Bruker APEX II (CCD area detector, Mo K$\alpha$, $\lambda$ = 0.7107 Å) at room temperature. Using the software packages of the corresponding diffractometer, data reduction, structure solution, and unitcell refinement were done and the suggested possible crystal structure of $PdBr_2$ comes-forth as monoclinic with space group C2/C. A bunch of as-grown crystals were taken for powder X-ray diffraction (PXRD) study without any further grinding. The PXRD data was collected using a Rigaku SmartLab (Cu $K_\alpha$ radiation, $\lambda$ = 1.5406 Å) diffractometer. The PXRD data was analyzed by Rietveld refinement, using the FULLPROF software. The simulated lattice parameters are given in Table 1.

\begin{table}[h!]
\caption{Lattice parameters of $PdBr_2$ (space group: C2/c)} 
\centering 
\begin{tabular}{c c c c c c c} 
\hline\hline 
Method & a & b & c & $\alpha$ & $\beta$ & $\gamma$ \\ [0.5ex] 
\hline 
SXRD & 13.04 & 3.996 & 6.662 & 90 & 102.597 & 90 \\
PXRD & 13.169 & 3.958 & 6.618 & 90 & 106.845 & 90 \\  [1ex] 
\hline 
\end{tabular}
\label{table:nonlin} 
\end{table}
%

The nanowires were prepared from the millimeter-sized as-grown brown and shiny bulk crystals by mechanical exfoliation using a scotch tape/homemade PDMS. The liquid exfoliation, dispersing the bulk crystals into toluene, was carried out  using a bath sonicator. In the dry transfer, the peeled off scotch-tapes were finally placed onto 285 nm
$SiO_2$/\textit{p}-Si substrates (Nova Electronic Materials) for future device fabrication and charaterization (transistor characteristics). Similarly, the liquid exfoliated sample was drop-casted onto suitable substrates and dried in ambient conditions. The nanowires were primarily examined under an optical microscope (OLYMPUS BX53M) to confirm their 1D structure. Transmission electron microscopic (TEM) images were obtained on a JEOL-JEM-F200 electron microscope with an accelerating voltage of 200 kV. Nanowires dispersed in toluene were dropped on a carbon-coated copper grid (Ted Pella) which was used for TEM imaging. The 1D structure of the exfoliated nanowires was also confirmed by scanning electron microscope (JEOL JSM-6010LA) and atomic force microscope (Asylum Research MFP-3D) on $Si/SiO_2$ substrates. The mentioned SEM quipped with the energy-dispersive spectroscopic (EDS) system  allows us to perform the chemical analysis of the individual flakes. X-ray photoelectron spectroscopy (XPS) experiments were performed in an Omicron electron spectrometer (model no. XM 1000). For characterizing the phonon modes, Raman spectroscopy was carried out using the Horiba T64000 raman spectrometer (excitation wavelength $\lambda$ = 532 nm with spot size $\sim$ 1$\mu$m)


Projection lithography setup was used to pattern (channel length varying from 5 - 20 $\mu$m) two-probe devices on predefined isolated nanowire, lying on $SiO_2$/Si substrate, followed by metallization of Cr/Au (10/70 nm). For sub-micron structures, electron beam lithography (Zeiss Sigma 300 with Raith Elphy Quantum) was utilized. Temperature-dependent electrical transport of fabricated devices were studied in a closed cycle cryostat (ARS-4HW) with Keithley 2450 source measure units (SMU). For the vertical transport, conducting tip (Pt coated Si tip from micromasch) is employed in the atomic force microscope ((Asylum Research MFP-3D) which maps current on individual  nanowires lying on conducting p-Si substrate. 


Electronic structure calculations are performed within density functional theory (DFT) framework using the Perdew-Burke-Ernzerhof (PBE) parametrization~\cite{Perdew1996} of the generalized gradient approximation (GGA) to the exchang-correlation (XC) functional, as implemented in QuantumATK package~\cite{QATK}; norm-conserving fully relativistic pseudopotentials of the type PseudoDojo-SO~\cite{QATK, Setten2018} for describing electron-core interactions; and the Pseudojojo (medium) numerical linear combination of atomic orbitals (LCAO) basis set~\cite{Setten2018}. As it is wellknown that the GGA-XC functional underestimate the bandgap, a more accurate hybrid Heyd-Scuseria-Ernzerhof (HSE06) functional~\cite{HSE06} is employed in the DFT calculations. The energy mesh cutoff for the real-space grid is chosen as 101 Hartree, and the k-point grid 4$\times$12$\times$7 is used for the self-consistent calculations. Electronic structure calculations are done using the experimental crystal structure with Triclinic lattice with $a= 13.04~\AA,~b=3.99~\AA, c=6.62\AA, \alpha= 90^{\circ}, \beta= 102^{\circ}, \gamma= 90^{\circ}$.
Electronic quantum transport calculations for two-terminal systems are performed by using density functional theory combined with nonequillibirum Green's function (DFT+NEGF) formalism using PBE-GGA XC functional as implemented in QuantumATK package~\cite{QATK}. Electronic transmission coefficient can be calculated as, $T(E)=\mathrm{Tr}[\Gamma_{\rm L}(E)G(E)\Gamma_{\rm R}(E)G^{\dagger}(E)]$, where $G(E)$ denotes the retarded Green’s function; broadening matrix $\Gamma_{\rm L,R}(E) = i[\Sigma_{\rm L,R}(E) - \Sigma_{\rm L,R}^{\dagger}(E)]$ are constructed from the self-energies $\Sigma_{\rm L,R}$ of left (L) or right (R) leads. The transmission coefficient $T(E)$ at energy $E$ was averaged over 101$\times$11 k-points perpendicular to the transport direction. Energy dependent conductance is defined as $G(E)=e^2/h[T(E)]$, where $e$ is the electron charge, $h$ is Planck constant.

\section{Results and Discussions}

From the structure calculation (see Figure 1(a)-(c)), it is evident that the unit cell of PdBr$_{\mathrm{2}}$ contains four Pd atoms and eight Br atoms where each Pd atom is coordinated to four Br atoms. This is due to the typical $d^8$ electronic configuration of $Pd ^{2+}$ complexes, which in general adopts
square-planar coordination \citep{structure}. In the case of  PdBr$_{\mathrm{2}}$, the square-planar units form an interconnected continuous chain which extends along the c-axis resulting in a  \textit{needle}-like structures. The \textit{truly}-1D chain with covalent bonds only in the chain direction,  are weakly bound in bundles by the vdW interaction which allows facile
single-chain exfoliation from the bulk crystal. In Figure 1(d), the optical micrograph of the representative single crystals is shown. Using a pointed tweezer or needle, flat fibres (see Figure 1(e)), stacked one after another, can easily be separated from the deformable parent crystal. 
The room temperature powder XRD (PXRD) data was recorded by taking a bunch of crystals together to confirm its phase purity. The PXRD pattern and refinement are shown in Figure 1(f). The lower value of the goodness of the fit parameter ($\chi^2 = 3.12$) suggests a good fit of the experimental data with the simulated one. The PXRD peaks at $14.13^{\circ}$ and $23.61^{\circ}$ correspond to (200) and (110) lattice planes of PdBr$_{\mathrm{2}}$ crystal system. Lattice parameters calculated from both SXRD and PXRD are given in Table 1. Single crystal refinement parameters are given in the supporting information.

Figure 2(a) depicts the TEM image of the 1D nature of the drop casted sample with variation in thicknesses/width, whereas a FESEM image of a single nanowire onto SiO$_{\mathrm{2}}$/Si substrate is given in Figure 2(b).
Figure S1(a) shows FESEM image of dropcasted sample onto SiO$_{\mathrm{2}}$/Si substrate where the individual nanowire with an average length $~200$ $\mu$m and mean width $\approx(40\pm$ 5) can be found (see Figure S1(b)). On the other hand, an optical microscopy image of micromechanically exfoliated PdBr$_{\mathrm{2}}$ nanowires on a PDMS substrate prepared by the dry transfer technique is shown in Figure S1(c) where broader nanowires (mean width $\approx(400\pm$ 20), see Figure S1(d)), but of similar length can be identified. Interestingly, in both cases, individual wires are branched out like a \textit{polytomy} from the root of a \textit{phylogenetic tree}. Note that the exfoliation of the polycrystalline chunk, left in the hot zone after the growth, showed the same quasi one-dimensional morphology when examined. For reaffirmation, atomic force microscopy (AFM) in tapping mode was performed on several drop casted nanowires (on SiO$_{\mathrm{2}}$/Si) to find the statistical distribution of  width/height of the fibres (see Figure 2(c)). Importantly, nodes for branching of fibres can be visualized from AFM micrograph. For the ease of device fabrication, one may choose the dropcasted samples with higher aspect ratio ($\sim$ 5000) over dry-transferred one. A representative FESEM image of a two probe single nanowire device fabricated by electron beam lithography (EBL) has been also shown in Figure 2(d).

The exact composition of the atomic elements and the atomic percentage ratio (Br:Pd = $\frac{64.6}{35.4}$ = 1.82) is obtained from EDS data. Moreover, in the HRTEM image (see Figure S2(a)), respective lattice spacings of 0.67 nm and 0.32 nm along (200), (202) plane is shown. Inter-planer spacing and their angles were matched using the crystallographic information file (CIF) extracted from the XRD refinement and VESTA software. From the selected area electron diffraction (SAED), as shown in in the Figure S2(b), first order bright spots in the FFT pattern can be assigned to different Bragg planes which is consistent with the d-spacing and angles in Figure S2(a). A STEM-HAADF image of a single nanoribbon has been shown in Figure S2(c). Also, homogeneity of the sample can be confirmed from the  HAADF-STEM image and correlated elemental mapping on a single nano-ribbon (shown in Figure S2(d) and (e)), where uniform spatial distribution of Pd and Br can be identified.

We investigated the optical absorbance characteristics (wavelength range between 300-800 nm) of the PdBr$_\mathrm{2}$ crystal in the solution phase (dispersed in Toluene) where a prominent peak is observed near 455 nm for the dispersed sample with proper baseline correction and the solvent artifact removal (see Figure S3(a)). The indirect energy bandgap (E$_g$) of the dispersed PdBr$_\mathrm{2}$ nanowires has been estimated as 0.87 eV by Tauc's plot method (see the inset of Figure S3(a)) \citep{tauc, tauc2}. Note that the observed abosrbance sprctra for different wire thicknesses indicated no evidence of indirect-to-direct band gap transition unlike single-atomic chain of V$_\mathrm{2}$Se$_\mathrm{9}$, as predicted in a recent report \citep{V2Se9} .

Before going for the microelectronic device fabrication, we examined the air stability of the freshly prepared (or transferred) exfoliated nano-ribbons with time by recording the morphological change using an optical micrograph. As a representative scenario, Figure S4 shows that the brown ribbon becomes completely black after 40 minutes of its exfoliation from the parent crystal which implies system's sensitivity to humidity and air exposure. Further, we used x-ray photoelectron spectroscopy (XPS) to determine the proper oxidation state of palladium for the bulk crystals. We probed the core-level 3d spectrum of Pd and Br (see Figure S5). XPS peaks located at 336.8 eV and 342.1 eV (with a separation of 5.3 eV) correspond to $Pd3d_{5/2}$ and $Pd3d_{3/2}$ spin-orbit components of PdO, respectively \cite{XPS1,XPS2,XPS3}. Other two peaks at 337.5 eV and 343.1 eV correspond to spin-orbit components of PdO$_\mathrm{2}$ \citep{XPS1}, indicating the surface oxidation of the material, as discussed above. Thus, surface doping may play a role in determining the proper electronic structure of the channel material as the device fabrication steps requires long hours during which the flakes might be degraded. The overlapping spin-orbit components of Br3d are given in Figure S5(b). $Br3d_{5/2}$ and $Br3d_{3/2}$ peaks located at 68.3 eV and 69.3 eV ($\delta\approx1 eV$) are also matching with other previous reports \citep{XPSBr}. 

For further stability studies of PdBr$_\mathrm{2}$ nanowires, micro-Raman scattering was performed (see Figure S3(b)). Note that minimum laser power (P$_{max}$ $\sim$ 100$\mu$W) on the surface of the flake was used in order to avoid photo-induced degradation as in the case of 2D black phosphorus \cite{BP}. For isolated nanowires, three distinct Raman active modes appearing at 163.5 $cm^{-1}$, 184.4 $cm^{-1}$ and 201.8 $cm^{-1}$ with a slight shift from the bulk counterpart can be observed. The small, but detectable shift (1-2 cm$^{-1}$) of the charateristic peaks with lowering thickness may be attributed to higher reactivity due to higher surface-to-volume ration in the case of 1D. 

Two/four-probe single nanowire devices of micrometer channel length (1 - 20 $\mu$m) were fabricated on \textit{p}-Si/SiO$_\mathrm{2}$ substrate using optical/e-beam lithography technique (see Experimental section for details). Low temperature resistivity measurement was performed in the temperature window 10-300K with $10 mV$ bias voltage. An increase of the resistance ($\frac{R_{10 K}}{R_{300 K}}$ $\sim$ 2) while cooling implies the semiconducting nature of the nanowire (For R \textit{vs.} T data, see Figure 3(a)). Corresponding AFM image of the representative device (nanowire height $\approx 55 nm$) is shown in the lower inset of the Figure 3(a). Effect of energetic disorder, \textit{i.e.} activation energy in an isolated wire, can be extracted from the $ln(R)$ \textit{vs.} $1/T$ graph (see Figure S6(a)). At higher temperature range (140-300 K), the charge transport is mainly dominated by the thermally activated band conduction where the variation of resistance with temperature follows Arrhenius relation, $R$(T)$\propto$ $exp(E_A/k_\mathrm{B}T)$, where $k_\mathrm{B}$ is the Boltzmann constant and $E_A$ is the activation energy of the charge carriers in the system \cite{Activation}. $E_A$ is calculated to be $\approx 10.8 meV$ for a single nanowire of given width. The calculated value of the activation energy is quite low compared to other reported 1D vdW systems, like TaSe$_\mathrm{3}$ ($E_A$ $\sim$ 0.9 eV), and can be attributed to the better extension of chain configuration under confinement in the case of PdBr$_\mathrm{2}$ which results in reduction of the morphological boundaries that hinders charge carrier transport \citep{noise}. 
\\
To understand the anomalous non-Arrhenius decay of the resistance ($R_{VRH}$(T)) at low temperature, Mott's variable range hopping (VRH) mechanism concerning phonon-assisted electron transport can be employed: $R_{VRH}(T)=R_{VRH,0} exp[(T_0/T)^{1/(1+d)}]$, where $T_0$ is the strength of the disorder and $d$ is the dimension (here, $d=1$) of the system \citep{Mott_VRH}. 
Note that $R_{VRH,0}$ has a temperature dependence as $R_{VRH,0} \propto T^{1/2}$.
$ln(RT^{-0.5})$ \textit{vs.} $\frac{1}{\sqrt{T}}$ is plotted in Figure S6(b).
Linear behaviour below 35 K (indicated by red in Figure S6(b)) is suggestive of the 1D Mott VRH dominated transport in the lower temperature regime. Though Schottky contact resistances at the two metal-semiconductor interfaces must play a role to determine the total resistance of the system, we considered only two terms as discussed above and neglected the contact resistance for simplicity \cite{SB, SB2}. The similar trends of resistivity curves were recorded for 8 devices. 

Room temperature current-voltage (I-V) characteristics were measured up to $\pm 0.5V$ (see the upper inset of Figure 3(a)). Nonlinear I-V characteristics were attributed due to the Schottky barrier formed at the nanowire and Cr/Au junctions. Figure 3(b) describes the I-V characteristics for six different nanowire devices till the \textit{break point}.  The distribution of the height and width of the nanowires considered for this study were in between 30-60 nm and 200-400 nm, respectively. The observed breakdown voltages vary in the range between 0.4 to 2.6 volt with a maximum current ($I_{max}$) 12 mA (see Figure 3(b)). Consequently, the current density ($J_b$) of a nanowire of typical dimension (30nm$\times$200nm) is calculated to be 200 $MA/cm^2$. Noticeably, this is two order-of-magnitude higher than that for copper, and one-order-of-magnitude higher than  quasi-1D vdW system $TaSe_3$ \cite{Breakdown1,Breakdown2}. It can be speculated from a very high current density that inside a single PdBr$_\mathrm{2}$ nanowire,there are comparatively less localized points of higher resistance, so less grain boundaries, impurities, or defects compared to the reported ones.  Importantly, capping the nanowires with insulating hexagonal boron nitride (h-BN) in a 1D-2D heterostructure \citep{thermalconductivityhBN1, thermalconductivityhBN2} or polymers (\textit{e.g.}PMMA) in polymer-vdW hybrid\cite{PMMA} may act as a protective layer of the nanowire to prevent its oxidation, and the device can achieve even higher breakdown current density. 

Furthermore, local electron transport measurements (perpendicular to the chain direction) were carried out on a single nanowire of different thicknesses (20-150 nm) utilizing the Pt-coated conducting tip in atomic force microscopy (CAFM) set-up. The nanowires were drop casted/exfoliated on \textit{p}-Si substrate. During the local current mapping, the tip was kept grounded and the substrate voltage was swept from 0 to 10 V. CAFM topography (see Figure 3(c)) and the corresponding bias voltage dependent current map ($V_b$ = +1 V) of a single nanowire (see (Figure 3(d)) can be correlated. Figure 3(e) depicts the height (red plot) and current (blue plot) variation following the line scan in Figure 3(c). The average height of the nanowire is $\approx 40$ nm with an average surface roughness of 1 nm. From the spatial current map of the nanowire and the contrast with Si, it is evident that the nanowire has higher conductivity ($\sim$ 100 times) than that of the substrate, as shown in Figure 3(d). Considering the minimum baseline current on the p-Si substrate, the maximum current ($I_{max}$) observed at the nanowire surface is 160 pA (see figure 3(e)). Reduction of current ($I_{max}$) with increasing height of the nanowire, in other words, higher number of resistive paths, can be identified in Figure 3(f). The reliability and the consistency of the result was checked on several nanowires.  The top-down current density for a maximum current of 185 pA (observed for a 24 nm thick wire) is calculated to be $\approx 15$ $A/cm^2$ which is 7 orders of magnitude lower than that of the current density value \textit{along-the chain} configuration, as discussed in the previous paragraph. Hence, electronic anisotropy which reflects the anisotropic band structure in the system can be predicted and studied further by using sophisticated experimental tools like angle resolved photoemission spectroscopy (ARPES).

To investigate the electronic structure and transport properties of $PdBr_2$, density functional theory
(DFT) combined with nonequillibirum Green’s function (NEGF) is used (see methodology for computational details).
Figure 4(a) shows the PBE bandstructure of bulk $PdBr_2$ exhibiting an indirect bandgap of 0.44 eV. However, more
accurate HSE-DFT calculation shows an enhancement of the indirect bandgap to 1.66 eV. The analysis of projected
density of states (DOS) reveals that both the conduction and valence band originate from Pd as well as Br atoms. Finally, we have carried out the
quantum transport calculations for two-terminal geometry in bulk pristine $PdBr_2$, where the transport direction is
along is either c-axis or b-axis. Conductance spectrum as shown in Figure 4(b), indicates the directional anisotropy of
eletronic transport, which is also observed in our experiment.
To further explore the bonding nature, we examined the electron localization function (ELF) \citep{ELF} of bulk PdBr$_\mathrm{2}$(see
Figure S7(a)). Almost no electron is localized between Pd and Br atoms, suggesting the typical ionic bonding, with Pd
atoms donating electrons to Br atoms. On the other hand, bulk $TiS_3$ which is another newly discovered 1D van der
Waals material, exhibit a localization of electrons between the Ti-S bonds – the localization is larger around the Ti
atoms in comparison to the S atoms – suggesting a polar covalent bonding of Ti-S bonds (see Figure S7(b)).

\section{Conclusions}
We report successful synthesis of a new truely 1D vdW material, PdBr$_\mathrm{2}$, which shows a maximum aspect ratio of 5000 when exfoliated. Moreover, lower value of
the current density in top-down geometry than that of the lateral configuration reveals the possible anisotropic band structure of the system. Also, high breakdown current density (200 MAcm$^{-2}$) indicates the use of PdBr$_2$ wire as a better interconnect than common metals in the semiconductor industry. These results provide a blueprint to explore this material with other vdW materials in $1D/2D$ or $1D/1D$ heterostructure based device applications. PdBr$_2$ nanowires and their magnetic hybrids, under spatial confinement (sub-micron or nanometer channel length), may provide ideal platform for studying exotic quantum states induced by Fermi surface instabilities and Kondo correlations.

\section{Acknowledgments}
The authors would like to thank Late Prof. Evan J. Reed (Standford University) for his inspiring work on computational materials science, especially on predicting the ``spectrum of 1D van der Waals materials'' in the context of this work. The authors would also acknowledge the fruitful discussions with Prof. Sudip Malik, Dr. Mintu Mondal and Prof. Abhishek Dey. A special thanks to Prof. Narayan Pradhan for the TEM facility. BD would also like to thank Dr. Sumit Kumar Dutta, Mr. Sourav Bera, Mr. Arghyadip Bhowmik, Dr. Rafikul Ali Saha, Mr. Souvik Banerjee, Mr. Parijat Biswas and Mr. Sayan Atta for the experimental help and structure/data analysis. SD would like to appreciate Dr. K. D. M. Rao's efforts as a faculty in-charge of e-beam lithography at IACS. BD is grateful to IACS for the fellowship. SM and TK are grateful to DST-INSPIRE for their fellowships. RP is grateful to CSIR for his fellowship. SD acknowledges the financial support from DST-SERB grant No. ECR/2017/002037 and CRG/2021/004334. SD also acknowledges support from the Central Scientific Service (CSS) and the Technical Research Centre (TRC), IACS, Kolkata.
 
\section{Keywords}
PdBr$_2$, 1D material, vdW material, Electrical Anisotropy

\bibliography{References}

\begin{thebibliography}{100}
\bibitem{Graphene}
Novoselov, K. S.; Geim, A. K.; Morozov, S. V.; Jiang, D.; Katsnelson, M. I.; Grigorieva, I. V.; Dubonos, S. V.; Firsov, A. A. Two-dimensional gas of massless Dirac fermions in
graphene. 
{\it Nature} 
{\bf 2005}, 438, 197–200.

\bibitem{MoS_2}
Radisavljevic, B.; Radenovic, A.; Brivio, J.; Giacometti, V.; Kis, A. Single-layer $MoS_2$ transistors. 
{\it Nat. Nanotech.} 
{\bf 2011}, 6(3), 147-150.

\bibitem{WS_2}
Zhao, W.; Ghorannevis, Z.; Chu, L.; Toh, M.; Kloc, C.; Tan, P.-H.; Eda, G. Evolution of Electronic Structure in Atomically Thin Sheets of $WS_2$ and $WSe_2$ .
{\it ACS Nano} 
{\bf 2013}, 7(1), 791–797.

\bibitem{MoTe_2}
Cho, S.; Kim, S.; Kim, J. H.; Zhao, J.; Seok, J.; Keum, D. H.; Baik, J.; Choe, D.-H.; Chang, K. J.; Suenaga, K.; Kim, S. W.; Lee, Y. H.; Yang, H. Phase patterning for ohmic
homojunction contact in $MoTe_2$.
{\it Science} 
{\bf 2015}, 349(6248), 625-628.

\bibitem{WTe_2}
Ali, M. N.; Xiong, J.; Flynn, S.; Tao, J.; Gibson, Q. D.; Schoop, L. M.; Liang, T.; Haldolaarachchige, N.; Hirschberger, M.; Ong, N. P.; Cava, R. J. Large, non-saturating magnetoresistance in $WTe_2$.
{\it Nature} 
{\bf 2014}, 514(7521), 205-208.

\bibitem{MoSe_2}
Kong, D.; Wang, H.; Cha, J. J.; Pasta, M.; Koski, K. J.; Yao, J.; Cui, Y. Synthesis of $MoS_2$ and $MoSe_2$ Films with Vertically Aligned Layers.
{\it Nano Lett.} 
{\bf 2013}, 13(3), 1341-1347.







\bibitem{QD_Science}
Klimov, V. I.; Mikhailovsky, A. A.; Xu, S.; Malko, A.; Hollingsworth, J. A.; Leatherdale, C. A.; Eisler, H.-J.; Bawendi, M. G. Optical Gain and Stimulated Emission in Nanocrystal Quantum Dots.
{\it Science} 
{\bf 2000}, 290(5490), 314-317.

\bibitem{Single Electron Transistor}
Klein, D. A.; Roth, R.; Lim, A. K. L.; Alivisatos, A. P.; McEuen, P. L. A single-electron transistor made from a cadmium selenide nanocrystal.
{\it Nature} 
{\bf 1997}, 389, 699–701.

\bibitem{Memory}
Pettersson, H.;  B\.{a}\.{a}th, L.;  Carlsson, N.; Seifert, W.; Samuelson, L. Case study of an InAs quantum dot memory:
Optical storing and deletion of charge.
{\it Appl. Phys. Lett.} 
{\bf 2001}, 79, 78.



\bibitem{TaSe3}
Barani, Z.; Kargar, F.; Ghafouri, Y.; Ghosh, S.; Godziszewski, K.; Baraghani, S.; Yashchyshyn, Y.; Cywiński, G.; Rumyantsev, S.; Salguero, T. T.; Balandin, A. A. Electrically Insulating Flexible Films with Quasi-1D van der 
Waals Fillers as Efficient Electromagnetic Shields in the 
GHz and Sub-THz Frequency Bands. 
{\it Adv. Mater.} 
{\bf 2021}, 31(11), 2007286.

\bibitem{TiS3}
land, J. O.; Barawi, M.; Biele, R.; Almazán, A.;    Clamagirand, J. M.; Ares, J. R.; Sánchez, C.; van der Zant, H. S. J.; Álvarez, J. V.; D’Agosta, R.; Ferrer, I. J.; Castellanos-Gomez, A. 
$TiS_3$  Transistors with Tailored Morphology and Electrical Properties.
{\it Adv. Mater.} 
{\bf 2015}, 27, 2595–2601.

\bibitem{Review1}
Balandin, A. A.; Kargar, F.; Salguero, T. T.; Lake, R. K. 
One-dimensional van der Waals quantum
materials.
{\it Materials Today} 
{\bf 2022}.

\bibitem{Review2}
Island, J. O.; Molina-Mendoza, A. J.; Barawi, M.; Biele, R.; Flores, E.; Clamagirand, J. M.; Ares, J. R.; Sánchez, C.; van der Zant, H. S. J.; D’Agosta, R.; Ferrer, I. J.; Castellanos-Gomez, A. 
Electronics and optoelectronics of quasi-1D
layered transition metal trichalcogenides.
{\it 2D Materials} 
{\bf 2017}, 4(2), 022003.

\bibitem{TaSe3SC}
Sambongi, T.; Yamamoto, M.; Tsutsumi, K.; Shiozaki, Y.; Yamaya, K.; Abe, Y. 
Superconductivity in one-dimensional $TaSe_3$.
{\it Journal of the Physical Society of Japan} 
{\bf 1977}, 42(4), 1421-1422.

\bibitem{NbSe3_1}
Slot, E.; Holst, M. A.; van der Zant, H. S. J.; Zaitsev-Zotov, S. V. 
One-Dimensional Conduction in Charge-Density-Wave Nanowires.
{\it Phys. Rev. Lett.} 
{\bf 2004}, 93, 176602.

\bibitem{NbSe3_2}
Slot, E.; Holst, M. A.; van der Zant, H. S. J.; Zaitsev-Zotov, S. V. 
Nanowires and Nanoribbons of Charge-Density-Wave Conductor $NbSe_3$.
{\it Nano Lett.} 
{\bf 2005}, 5(2), 397–401.

\bibitem{TiS3MIM}
Randle, M.; Lipatov, A.; Kumar, A.; Kwan, C.-P.; Nathawat, J.; Barut, B.; Yin, S.; He, K.; Arabchigavkani, N.; Dixit, R.; Komesu, T.; Avila, J.; Asensio, M. C.; Dowben, P. A.; Sinitskii, A.; Singisetti, U.; Bird, J. P. 
Gate-Controlled Metal–Insulator Transition in $TiS_3$ Nanowire Field-Effect Transistors.
{\it ACS Nano} 
{\bf 2019}, 13(1), 803–811.


\bibitem{TaS3}
Li, W.; Yang, L.; Wang. J.; Xiang, B.; Yu, Y. 
Three-Dimensionally Interconnected $TaS_3$ Nanowire Network as
Anode for High-Performance Flexible Li-Ion Battery.
{\it ACS Appl. Mater. Interfaces} 
{\bf 2015}, 7, 5629-5633.

\bibitem{TiS3_2}
Island, J. O.; Buscema, M.; Barawi, M.; Clamagirand, J. M.; Ares, J. R.; Sánchez, C.; Ferrer, I. J.; Steele, G. A.; van der Zant, H. S. J.; Castellanos-Gomez, A. Ultrahigh Photoresponse of Few-Layer $TiS_3$ Nanoribbon Transistors.
{\it Adv. Optical Mater.} 
{\bf 2014}, 2, 641–645.

\bibitem{ZrTe3}
Geremew, A.; Bloodgood, M. A.; Aytan, E.; Woo, B. W. K.; Corber, S. R.; Liu, G.; Bozhilov, K.; Salguero, T. T.; Rumyantsev, S.; Rao, M. P.; Balandin, A. A. Current Carrying Capacity of Quasi-1D $ZrTe_3$
Van Der Waals Nanoribbons.
{\it IEEE Electron Device Letters} 
{\bf 2018}, 39(5), 735-738.


\bibitem{Halide1}
Perfetti, L.; Berger, H.; Reginelli, A.; Degiorgi, L.; Höchst, H.; Voit, J.; Margaritondo, G.; Grioni, M. Spectroscopic Indications of Polaronic Carriers in in the Quasi-One-Dimensional Conductor $(TaSe_4)_2I$. 
{\it Phys. Rev. Lett.} 
{\bf 2001}, 87(21), 216404.

\bibitem{arnab}
Bera, A.; Gayen, S.; Mondal, S.; Pal, R.; Pal, B.; Vasdev, A; Howlader, S.; Jana, M.; Maiti, T.; Saha, R. A.; Das, B.; Satpati, B.; Pal, A. N.; Mandal, P.; Sheet, G.; Mondal, M. Superconductivity coexisting with ferromagnetism in a quasi-one dimensional non-centrosymmetric $(TaSe_4)_3I$. 
{\it arXiv preprint arXiv:2111.14525 (2021)}.


\bibitem{Halide2}
Vescoli, V.; Zwick, F.; Voit, J.; Berger, H.; Zacchigna, M.; Degiorgi, L.; Grioni, M.; Grüner, G. Dynamical Properties of the One-Dimensional Band Insulator $(NbSe_4)_3I$. 
{\it Phys. Rev. Lett.} 
{\bf 2000}, 84(6), 1272.


\bibitem{Halide4}
Autès, G.;  Isaeva, A.; Moreschini, L.; Johannsen, J. C; Pisoni, A.; Mori, R.; Zhang, W.; Filatova. T. G.; Kuznetsov, A. N.; Forró, L.; den Broek, W. V.; Kim, Y.; Kim, K. S.;  Lanzara, A.; Denlinger, J. D.; Rotenberg, E.; Bostwick, A.; Grioni, M.; Yazyev, O. V. A novel quasi-one-dimensional topological
insulator in bismuth iodide $\beta-Bi_4I_4$. 
{\it Nat. Mater.} 
{\bf 2016}, 15, 154–158.

\bibitem{Halide5}
Liu, Y.; Chen, R.; Zhang, Z.; Bockrath, M.; Lau, C. N.; Zhou, Y.-F.; Yoon, C.; Li, S.; Liu, X.; Dhale, N.; Lv, B.; Zhang, F.; Watanabe, K.; Taniguchi, T.; Huang, J.; Yi, M.; Oh, J. S.; Birgeneau, R. J. Gate-Tunable Transport in Quasi-One-Dimensional $\alpha-Bi4I4$ Field Effect Transistors. 
{\it Nano. Lett.} 
{\bf 2022}, 22(3), 1151–1158.

\bibitem{Reed}
Zhu, Y.; Rehn, D.A.; Antoniuk, E. R.; Cheon, G.; Freitas, R.; Krishnapriyan, A.; Reed, E. J. Spectrum of Exfoliable 1D van der Waals
Molecular Wires and Their Electronic
Properties. 
{\it ACS Nano} 
{\bf 2021}, 15, 9851-9859.
\bibitem{CrCl3}
McGuire, M. A.; Clark, G.; KC, S.; Chance, W. M.;   Jr., G. E. J,; Cooper, V. R.; Xu, X.; Sales, B. C. Magnetic behavior and spin-lattice coupling in cleavable van der Waals layered $CrCl_3$ crystals. 
{\it Phys. Rev. Mater.} 
{\bf 2017}, 1, 014001. 

\bibitem{RuCl3}
May, A. F.; Yan, J.; McGuire, M. A. A practical guide for crystal growth of van
der Waals layered materials. 
{\it J. Appl. Phys.} 
{\bf 2020}, 128, 051101. 

\bibitem{structure}
Zhao, X. W.; Yang, Z.; Guo, J. T.; Hu, G. C.; Yue, W. W., Yuan, X. B.; Ren, J. F. Tuning electronic and optical 
properties of monolayer PdSe2 by 
introducing defects: frst-principles 
calculations. 
{\it Sci. Rep.} 
{\bf 2020}, 10(1), 1-8. 



\bibitem{tauc}
Maku\l{}a, P.; Pacia, M.; Macyk, W. How To Correctly Determine the Band Gap Energy of Modified Semiconductor Photocatalysts Based on UV-Vis spectra. 
{\it J. Phys. Chem. Lett.} 
{\bf 2018}, 9, 6814-6817. 

\bibitem{tauc2}
Yu, X.; Pr\'{e}vot, M. S.; Guijarro, N; Sivula, K. Self-assembled 2D WSe2 thin films
for photoelectrochemical hydrogen production. 
{\it Nat. Commun.} 
{\bf 2015}, 6(1), 1-8. 

\bibitem{V2Se9}
Le, W.-G.; Chae, S.; Chung, Y. K.; Yoon, W.-S.; Choi, J.-Y.; Huh, J. Indirect-To-Direct Band Gap Transition of One-Dimensional $V_2Se_9$:
Theoretical Study with Dispersion Energy Correction. 
{\it ACS Omega} 
{\bf 2019}, 4, 18392–18397. 

\bibitem{XPS1}
Guerrero-Ortega, L. P. A.; Ram\'{i}rez-Meneses, E.; Cabrera-Sierra, R.; Palacios-Romero, L. M.; Philippot, K.; Santiago-Ram\'{i}rez, C. R.; Lartundo-Rojas, L,; Manzo-Robledo, A. Pd and Pd@PdO core–shell nanoparticles supported
on Vulcan carbon XC-72R: comparison
of electroactivity for methanol electro-oxidation
reaction. 
{\it J. Mater. Sci.} 
{\bf 2019}, 54, 13694–13714. 

\bibitem{XPS2}
Voogt, E. H.; Mens, J. M.; Gijzeman, O. L. J.; Geus, J. W. XPS analysis of palladium oxide layers and particles. 
{\it Surf. Sci.} 
{\bf 1996}, 350, 21-31. 


\bibitem{XPS3}
Pillo, T.; Zimmermann, R.; Steiner, P.; H\"{u}fner, S. 
The electronic structure of PdO found by
photoemission (UPS and XPS) and inverse
photoemission (BIS). 
{\it J. Phys.: Condens. Matter} 
{\bf 1997}, 9, 3987–3999.
 
 
\bibitem{XPSBr}
Papirer, E.; Lacroix, R.; Donnet, J.-B.; Nanse, G.; Fioux, P. 
XPS Study of the halogenation of carbon black-part 1. Bromination. 
{\it Carbon} 
{\bf 1994}, 32, 1341-1358.

\bibitem{BP}
Ahmed, T.; Balendhran, S.; Karim, M. N.; Mayes, E. L.; Field, M. R.; Ramanathan, R.; Singh, M.; Bansal, V.; Sriram, S.; Bhaskaran, M.; Walia, S. Degradation of black phosphorus is contingent on UV–blue light exposure. 
{\it npj 2D Materials and Applications} 
{\bf 2017}, 1, 18.


\bibitem{Activation}
Radisavljevic, B.; Kis, A.
Mobility engineering and a metal–insulator
transition in monolayer $MoS_2$ 
{\it Nat. Mater.} 
{\bf 2013}, 12(9), 815-820.

\bibitem{noise}
Liu, G.; Rumyantsev, S.; Bloodgood, M. A.; Salguero, T. T.; Shur, M.; Balandin, A. A. Low-frequency electronic noise in quasi-1D $TaSe_3$ van der Waals nanowires. 
{\it Nano Lett.} 
{\bf 2017}, 17, 377-383.

\bibitem{VRH1}
Mott, N. F. 
Conduction in glasses containing transition metal ions. 
{\it Non-Cryst. Solids} 
{\bf 1968}, 1(1), 1-17.


\bibitem{VRH2}
Efros, A. L.; Shklovskii, B. I.;
Coulomb gap and low temperature conductivity of
disordered systems. 
{\it J. Phys. C: Solid State Phys.} 
{\bf 1975}, 8(4), L49.

\bibitem{VRH3}
Kumar, R.; Khare, N. Temperature dependence of conduction mechanism of ZnO
and Co-doped ZnO thin films. 
{\it Thin Solid Films} 
{\bf 2008}, 516, 1302–1307.

\bibitem{VRH4}
Paasch, G.; Lindner, T.; Scheinert, S. Variable range hopping as possible origin of a universal relation
between conductivity and mobility in disordered
organic semiconductors. 
{\it Synthetic Metals} 
{\bf 2002}, 132, 97-104.

\bibitem{VRH5}
Lin, T. T.; Young, S. L.;  Kung, C. Y.; Chen, H. Z.; Kao, M. C.; Chang, M. C.; Ou, C. R. Variable-range hopping and thermal activation conduction of Y-doped ZnO nanocrystalline films.
{\it IEEE Transactions on Nanotechnology} 
{\bf 2013}, 13(3), 425-430.

\bibitem{Mott_VRH}
Mott, N. F. Conduction in non-crystalline materials. 
{\it Nat. Mater.} 
{\bf 1969}, 19 (160), 835–852.


\bibitem{SB}
Allain, A.; Kang, J.; Banerjee, K.; Kis, A. 
Electrical contacts to two-dimensional 
semiconductors. 
{\it Philosophical Magazine. Informa UK Limited.} 
{\bf 2015}, 14(12), 1195-1205.

\bibitem{SB2}
Schroder, D. K.; Meier, D. L.
Solar cell contact resistance—A review. 
{\it IEEE Transactions on electron devices} 
{\bf 1984}, 31(5), 637-647.


\bibitem{Breakdown1}
Empante, T. A.; Martinez, A.; Wurch, M.; Zhu, Y.; Geremew, A. K.; Yamaguchi, K.; Isarraraz, M.; Rumyantsev, S.; Reed, E. J.; Balandin, A. A.; Bartels, L. 
Low Resistivity and High Breakdown Current Density of 10 nm Diameter van der Waals $TaSe_3$ Nanowires by Chemical Vapor Deposition.
{\it Nano. Lett.} 
{\bf 2019}, 19(7), 4355–4361.

\bibitem{Breakdown2}
Stolyarov, M. A.; Liu, G.; Bloodgood, M. A.; Aytan, E.; Jiang, C.; Samnakay, R.; Salguero, T. T.; Nika, D. L.; Rumyantsev, S. L.; Shur, M. S.; Bozhilov, K. N.; Balandin, A. A. 
Breakdown current density in h-BN-capped
quasi-1D TaSe3 metallic nanowires: prospects of
interconnect applications.
{\it Nanoscale} 
{\bf 2016}, 8, 15774-15782.

\bibitem{thermalconductivityhBN1}
Yuan, C.; Li, J.; Lindsay, L.; Cherns, D.; Pomeroy, J. W,; Liu, S.; Edgar, J. H.;  Kuball, M.
Modulating the thermal conductivity in hexagonal
boron nitride via controlled boron isotope
concentration.
{\it Communications Physics} 
{\bf 2019}, 2, 43.

\bibitem{thermalconductivityhBN2}
Cai, Q.;  Scullion, D.; Gan, W.; Falin, A.; Zhang, S.; Watanabe, K.; Taniguchi, T.; Chen, Y.; Santos, E. J. G.; Li, L. H.
High thermal conductivity of high-quality monolayer
boron nitride and its thermal expansion.
{\it Sci. Adv.} 
{\bf 2019}, 5(6), eaav0129.


\bibitem{PMMA}
Früh, A.;  Egelhaaf, H.-J.;  Hintz, H.;  Quinones, D.; Brabec, C. J.; Peisert, H.;  Chassé, T.
PMMA as an effective protection layer against the oxidation of P3HT
and MDMO-PPV by ozone.
{\it Journal of Materials Research} 
{\bf 2018}, 33(13)), 1-11.

\bibitem{Zhu2021} Zhu, Y.; Rehn, D. A.; Antoniuk, E. R.; Cheon, G.; Freitas, R.; Krishnapriyan, A.; Reed, E. J. Spectrum of Exfoliable 1D van der Waals Molecular Wires and Their Electronic Properties. 
{\it ACS Nano} 
{\bf 2021}, 15, 9851-9859.

\bibitem{Perdew1996} Perdew, J. P.; Burke, K.; Ernzerhof, M. Generalized Gradient Approximation Made Simple. 
{\it Phys. Rev. Lett.} 
{\bf 1996} 77, 3865-3868. 

\bibitem{QATK} Smidstrup, S.; Markussen, T.; Vancraeyveld, P.; Wellendorff, J.; Schneider, J.; Gunst, T.; Verstiche, B.; Stradi, D.; Khomyakov, P. A.; Vej-Hansen, U. G.; Lee, M.-E.; Chill, S. T.; Rasmussen, F.; Penazzi, G.; Corsetti, F.; Ojanperä, A.; Jensen, K.; Palsgaard, M. L. N.; Martinez, U.; Blom, A.; Brandbyge, M.; Stokbro, K. QuantumATK: an integrated platform of electronic and atomic-scale modeling tools. 
{\it J. Phys.: Condens. Matter} 
{\bf 2019}, 32, 015901.

\bibitem{Setten2018} van Setten, M.J.; Giantomassi, M.; Bousquet, E.; Verstraete, M. J.; Hamann, D. R.; Gonze, X.; Rignanese, G. M. The PseudoDojo: Training and grading a 85 element optimized norm-conserving pseudopotential table.
{\it Comp. Phys. Comm.} 
{\bf 2018}, 226, 226.

\bibitem{HSE06} Heyd, J.; Scuseria, G. E.; Ernzerhof, M. Hybrid functionals based on a screened Coulomb potential.
{\it J. Chem. Phys.} 
{\bf 2003}, 118, 8207.

\bibitem{ELF} Becke, A. D.; Edgecombe, K. E. A simple measure of electron localization in atomic and molecular systems. 
{\it J. Chem. Phys.} 
{\bf 1990}, 92, 5397.
 

\end{thebibliography}

\begin{figure}[ht]
\centerline{\includegraphics[scale=0.7, clip]{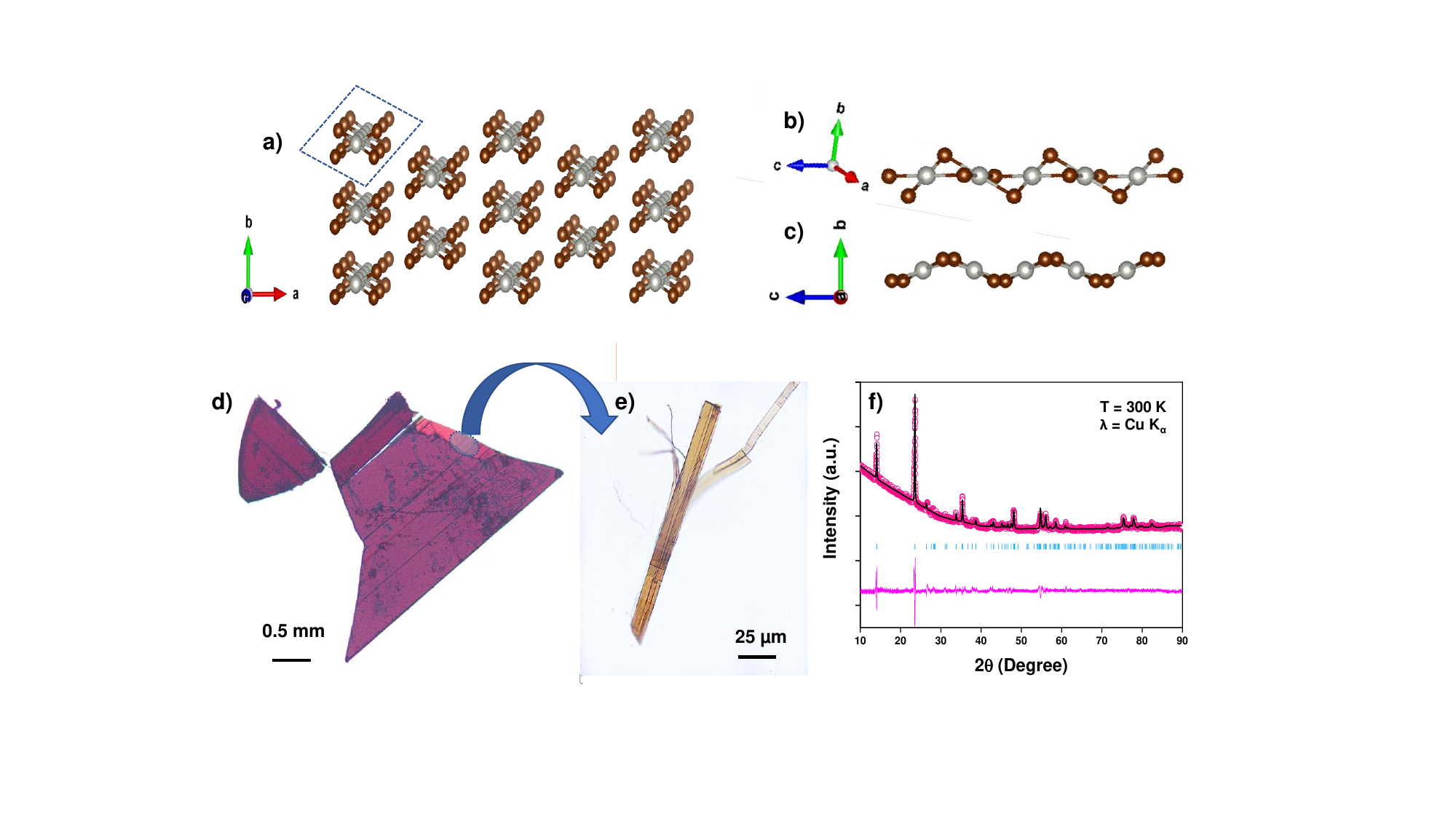}}
\caption{(a) Structure of $PdBr_2$ crystal system (space group: C2/c). The region identified by the blue dashed square is a true one dimensional wire. (b) and (c) are a true 1D wire shown along different perspectives. (d) An image of an as-grown $PdBr_2$ single crystal with a scale bar attached. (e) Fibre-like structure of an exfoliated quasi 1D $PdBr_2$ on PDMS substrate. (f) Powder XRD pattern with proper fitting of the experimental data with the simulated data at room temperature using Cu $K_{\alpha}$ radiation.
\label{structure}}
\end{figure} 
\begin{figure}[ht]
\centerline{\includegraphics[scale=.9, clip]{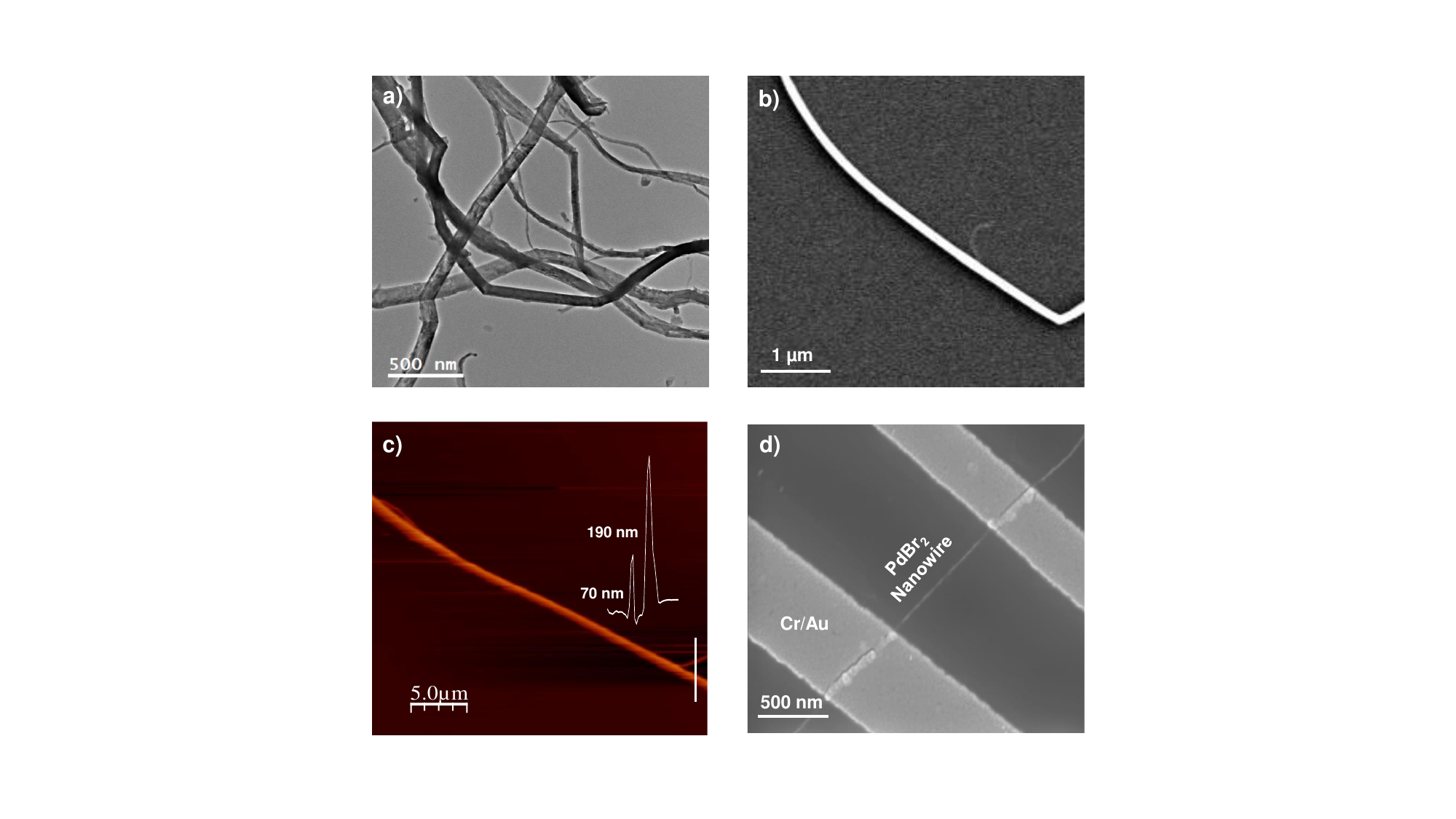}}
\caption{(a) TEM images of the quasi 1D $PdBr_2$ nanowire/nanoribbon. (b) FESEM image of a single $PdBr_2$ nanowire prepared by liquid exfoliation of the crystals in tolune dropped on Si/$SiO_2$ substrate. (c) Tapping mode atomic force microscopy (AFM) of drop casted nanowires (on SiO$_{\mathrm{2}}$/Si) substrate. Nodes for branching of fibres can be visualized from the image. The inset shows the height profile of two nanowires according to the white line. (d) Representative image of a two probe nanowire device fabricated by electron beam lithography (EBL) on $Si/SiO_2$
\label{device}}
\end{figure}

\begin{figure}[ht]
\centerline{\includegraphics[scale=.6, clip]{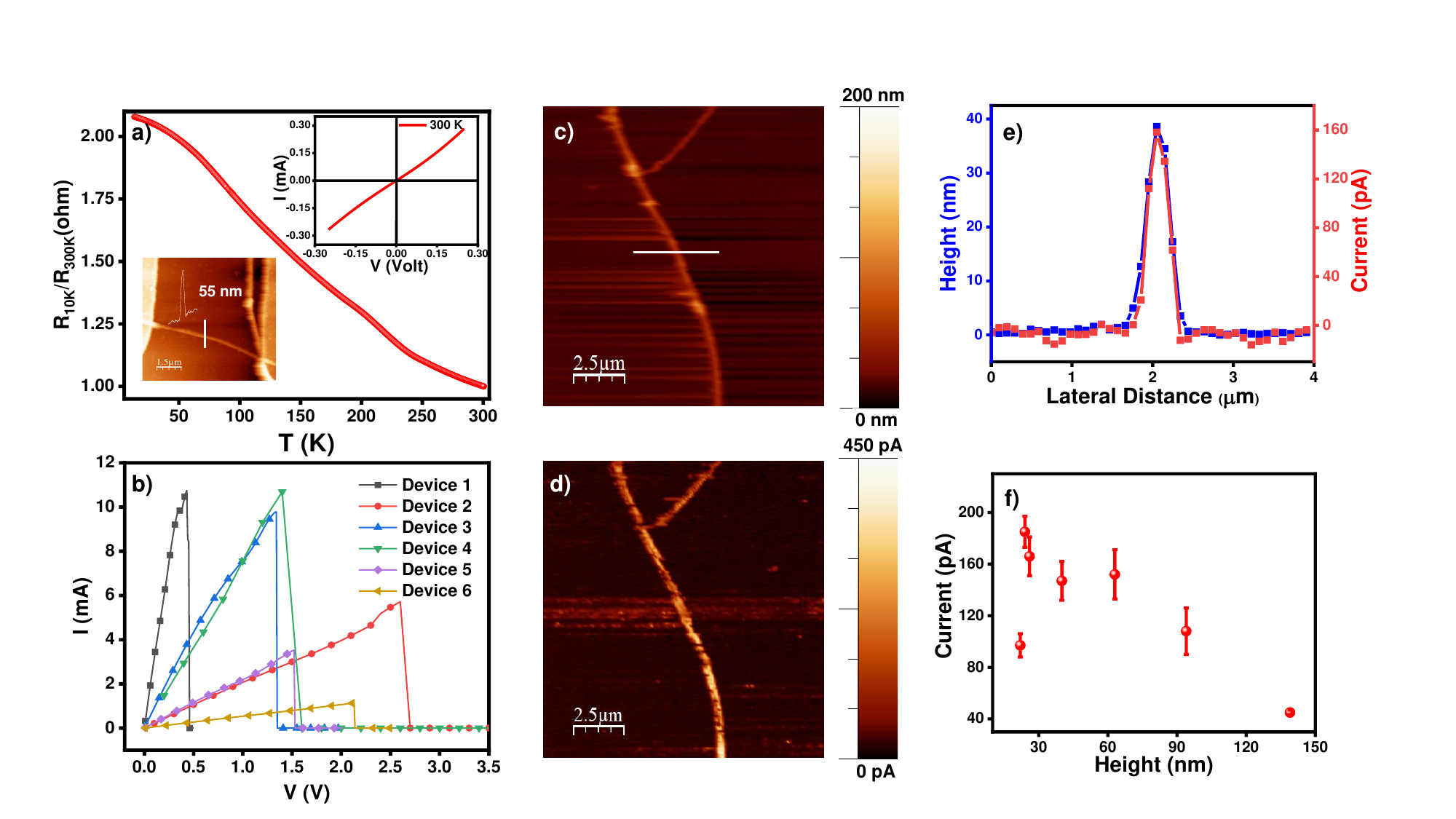}}
\caption{(a) Two probe temperature dependent resistance data indicating the semiconducting property of a single nanowire device. Lower inset: a representative AFM image of a single nanowire device. Upper inset: room temperature current-voltage characteristics of the device as shown. (b) I-V curve for six different devices until their complete breakdown. The sudden jumps in current correspond to the breakdown voltages (0.4-2.6 V) of the wires. (c) Topographical image and (d) correlated current map on a single nanowire on a p-Si substrate at applied bias +1 volt using CAFM. (e) Height (blue) and current (red) profile along the line cut drawn on (c) implies the nanowire has higher conductivity than the conductivity of the substrate. (f) Top-down current \textit{vs.} height data of several nanowires show a lower value of current for a thicker wire. 
\label{Transport}}
\end{figure}
\begin{figure}[ht]
\centerline{\includegraphics[scale=.6, clip]{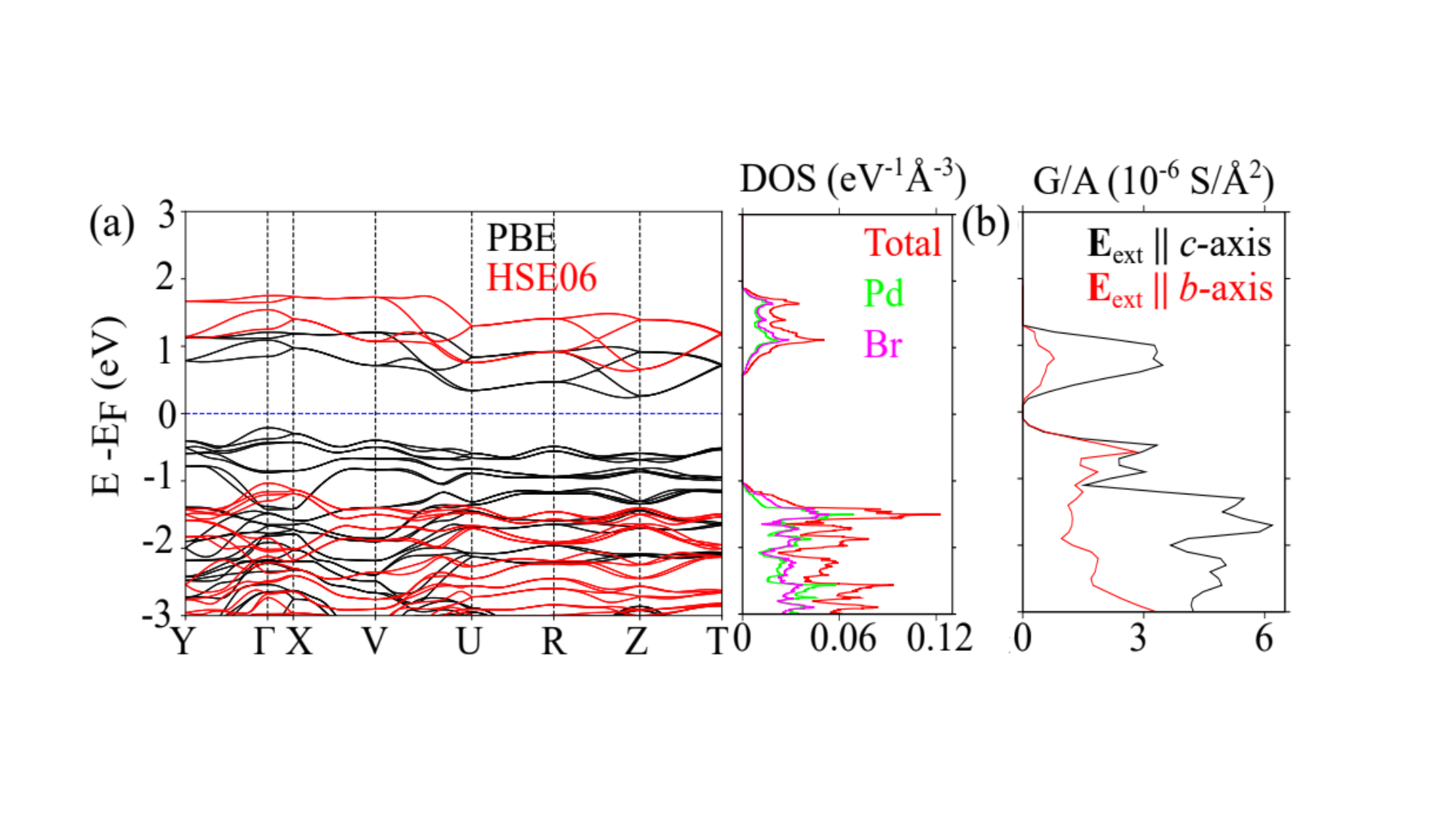}}
\caption{(a) Band structure (left) and density of states, DOS (right) of a bulk $PdBr_2$ with a Triclinic crystal structure. The
black and red lines denote the bandstrucures calculated using PBE and HSE06 XC-functional, respectively. Here, the DOS
corresponds to the HSE06 bandstructure. The green and magenta curves indicate the total DOS (red curve) projected on
the Pd and Br atoms, respectively. (b) Theoretically
calculated energy-dependent conductance G(E) (per unit area A) at infinitesimal small bias, when an external electric field $E_{ext}$
is applied along c-axis (black line) and b-axis (red line).
\label{DFT}}
\end{figure}

\newpage

\end {document}


\title{Supporting information for \\``Ultrahigh breakdown current density of van der Waals One Dimensional PdBr$_{\mathrm{2}}$"}

\author{Bikash Das}
\affiliation{School of Physical Sciences, Indian Association for the Cultivation of Science, 2A \& B
Raja S. C. Mullick Road, Jadavpur, Kolkata - 700032, India}

\author{Kapildeb Dolui}
\affiliation{Lomare Technolgies Limited, 6 London Street, London EC3R 7LP, United Kingdom}

\author{Rahul Paramanik}
\affiliation{School of Physical Sciences, Indian Association for the Cultivation of Science, 2A \& B
Raja S. C. Mullick Road, Jadavpur, Kolkata - 700032, India}

\author{Tanima Kundu}
\affiliation{School of Physical Sciences, Indian Association for the Cultivation of Science, 2A \& B
Raja S. C. Mullick Road, Jadavpur, Kolkata - 700032, India}

\author{Sujan Maity}
\affiliation{School of Physical Sciences, Indian Association for the Cultivation of Science, 2A \& B
Raja S. C. Mullick Road, Jadavpur, Kolkata - 700032, India}

\author{Anudeepa Ghosh}
\affiliation{School of Physical Sciences, Indian Association for the Cultivation of Science, 2A \& B
Raja S. C. Mullick Road, Jadavpur, Kolkata - 700032, India}

\author{Mainak Palit}
\affiliation{School of Physical Sciences, Indian Association for the Cultivation of Science, 2A \& B
Raja S. C. Mullick Road, Jadavpur, Kolkata - 700032, India}

\author{Subhadeep Datta}
\affiliation{School of Physical Sciences, Indian Association for the Cultivation of Science, 2A \& B
Raja S. C. Mullick Road, Jadavpur, Kolkata - 700032, India}

\email{sspsdd@iacs.res.in}
\maketitle



\begin{table}[h!]
\caption{Single Crystal refinement parameters of $PdBr_2$} 
\centering 
\begin{tabular}{c c} 
\hline 
CCDC No.	& 2205108 \\
Empirical formula	& $PdBr_2$  \\  
Formula weight	& 266.22 \\
Temperature/K	& 273.15\\
Crystal system	& 	monoclinic\\
Space group	& 	C2/c\\
a/Å	& 13.04(2)\\
b/Å	& 3.996(6)\\
c/Å	& 6.662(10)\\
$\alpha$/°	& 90\\
$\beta/$°	& 102.597(18)\\
$\gamma/$°	& 90\\
Volume/$Å^3$	& 338.9(9)\\
Z	& 4\\
$\rho_{calc}$ $g/cm^3$	& 5.218\\
$\mu/mm^{‑1}$	& 28.755\\
F(000)	& 464.0\\	
Crystal size/$mm^3$	& 0.21 × 0.13 × 0.12\\
Radiation	& MoK$\alpha$ ($\lambda = 0.71073$)\\
2$\theta$ range for data collection/°	& 10.698 to 50.752\\
Index ranges	& $-12 \leq h \leq 15$, $-4 \leq k \leq 4$, $-8 \leq l \leq 6$\\
Reflections collected	& 	894\\
Independent reflections	& 299 [$R_{int} = 0.0644$, $R_{sigma} = 0.0727$]\\
Data/restraints/parameters	& 299/0/16\\
Goodness-of-fit on $F^2$	& 1.182\\
Final R indexes [$I>=2\sigma (I)$]	& $R_1 = 0.0988$, $wR_2 = 0.2651$\\	
Final R indexes [all data] & $R_1 = 0.1027$, $wR_2 = 0.2692$\\
Largest diff. peak/hole / e $Å^{-3}$	& 4.35/-3.74\\
\hline 
\end{tabular}
\label{table:nonlin} 
\end{table}

 \begin{figure}[ht]
\centerline{\includegraphics[scale=0.8, clip]{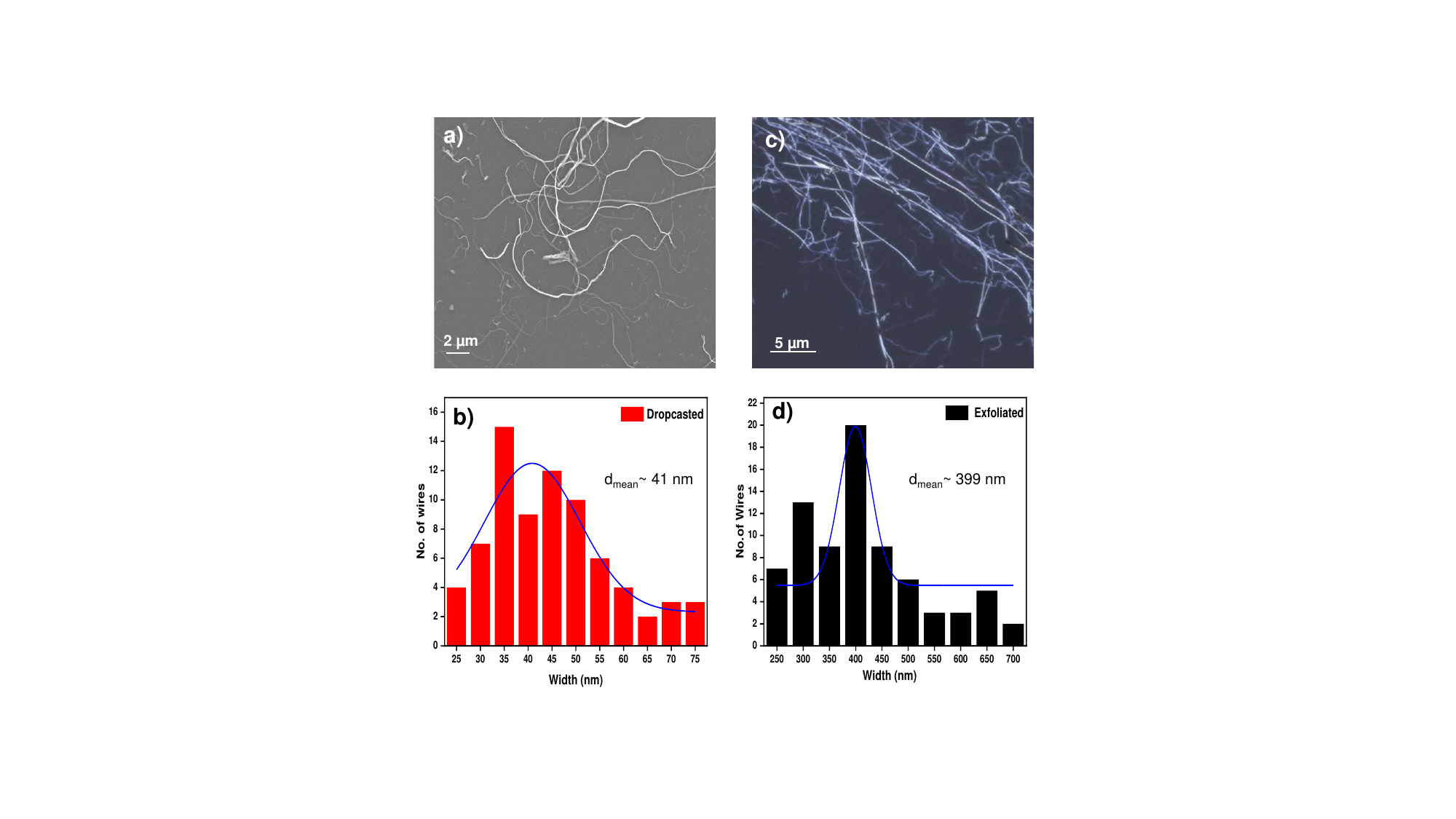}}
\renewcommand{\thefigure}{S\arabic{figure}}
\caption{(a) FESEM image of $PdBr_2$ nanowires prepared by liquid exfoliation of the crystals in tolune dropped on Si/$SiO_2$ substrate. (b) The number of wire \textit{vs.} width of the nanowires histogram corresponding to Figure (a). The mean width of the nanowires was calulated to be $41\pm5$ nm. (c) Optical microscope image of mechanically exfoliated $PdBr_2$ nanowires on PDMS substrate. (d) Number of wire \textit{vs.} width of the nanowires histogram corresponding to Figure (b). The mean width of the nanowires is $399\pm20$ nm here.\label{stat}}
\end{figure}

\begin{figure}[ht]
\centerline{\includegraphics[scale=0.7, clip]{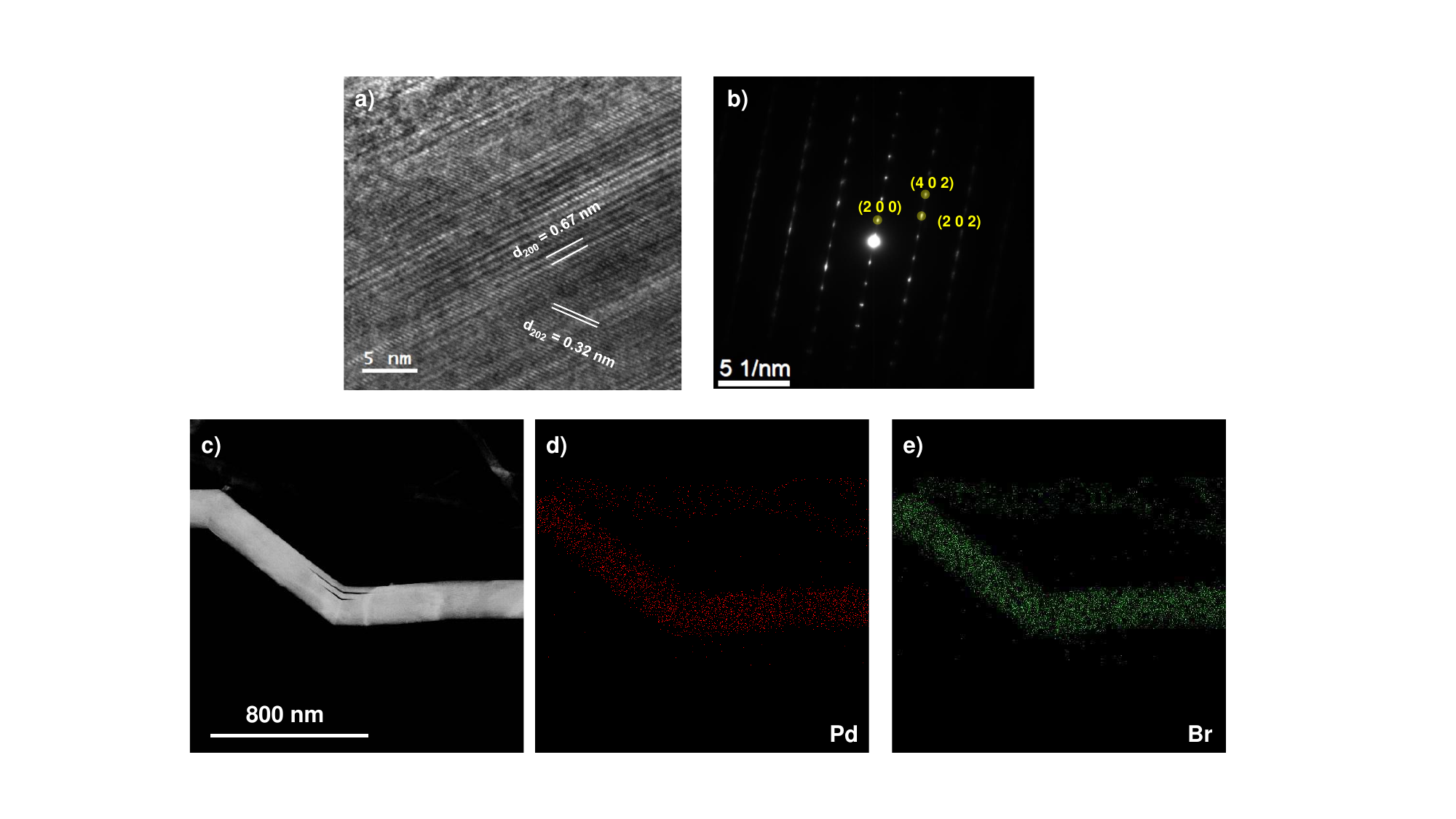}}
\renewcommand{\thefigure}{S\arabic{figure}}
\caption{(a)HRTEM data showing (200) and (202) lattice planes with their calculated lattice spacing of 0.67 and 0.32 nm, respectively. (b) SAED pattern of the $PdBr_2$ nanowire showing different bright spots associated with different lattice planes. (200), (202) and (402) planes are indicated in the data. HAADF-STEM image (c) and correlated elemental mapping (d) and (e) for Pd and Br, respectively performed on a single $PdBr_2$ nano-ribbon, where uniform spatial distribution of Pd and
Br can be identified.\label{TEM}}
\end{figure}

\begin{figure}[ht]
\centerline{\includegraphics[scale=0.7, clip]{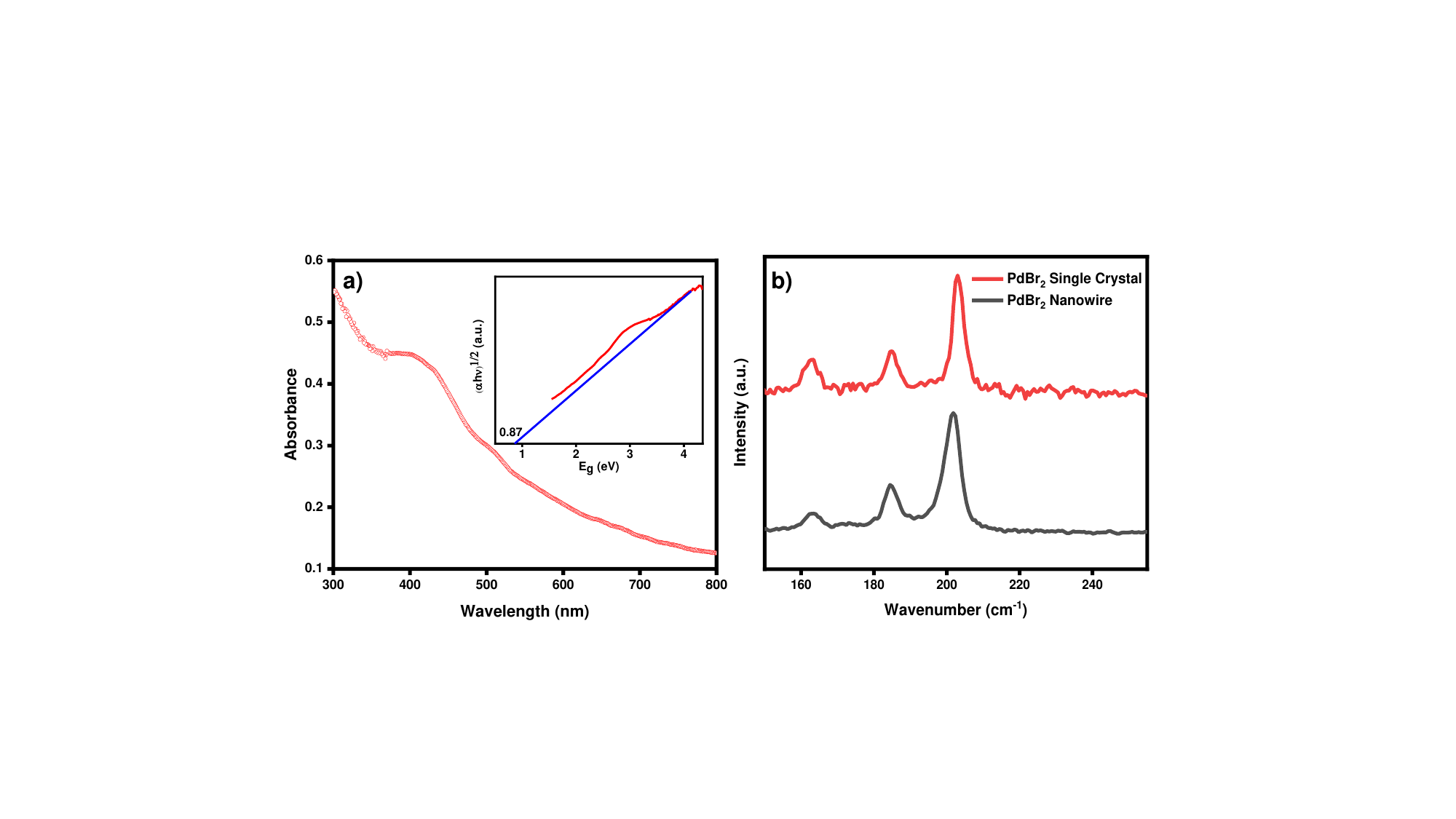}}
\renewcommand{\thefigure}{S\arabic{figure}}
\caption{(a) The optical absorbance characteristics of the $PdBr_2$ crystal in the solution phase (dispersed in Toluene) where a prominent
peak is observed near 455 nm for the dispersed sample with proper baseline correction. Inset: the indirect energy bandgap ($E_g$) of the
dispersed PdBr2 nanowires has been estimated as 0.87 eV by Tauc’s plot method. (b) Micro-Raman spectra for $PdBr_2$ single crystal and exfoliated nanowire. For both, three distinct Raman active modes were appeared at 163.5 $cm^{-1}$, 184.4 $cm^{-1}$ and 201.8 $cm^{-1}$.
\label{UV_Raman}}
\end{figure} 

\begin{figure}[ht]
\centerline{\includegraphics[scale=0.7, clip]{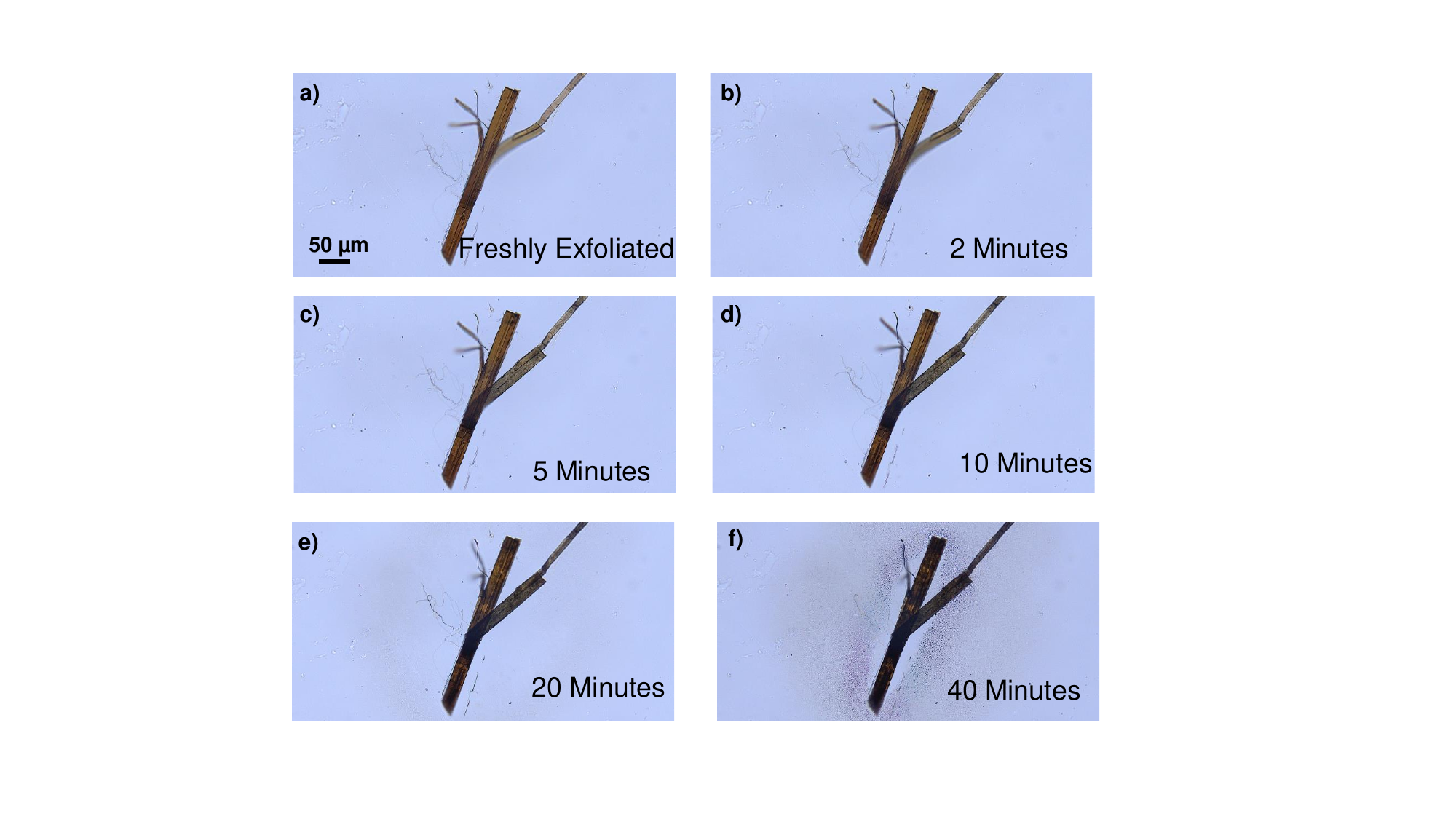}}
\renewcommand{\thefigure}{S\arabic{figure}}
\caption{Optical micrope images of the an exfolited nano-ribbon over time. The brown ribbon becomes completely black after
40 minutes of its exfoliation from the parent crystal.
\label{Degrade}}
\end{figure} 

\begin{figure}[ht]
\centerline{\includegraphics[scale=0.7, clip]{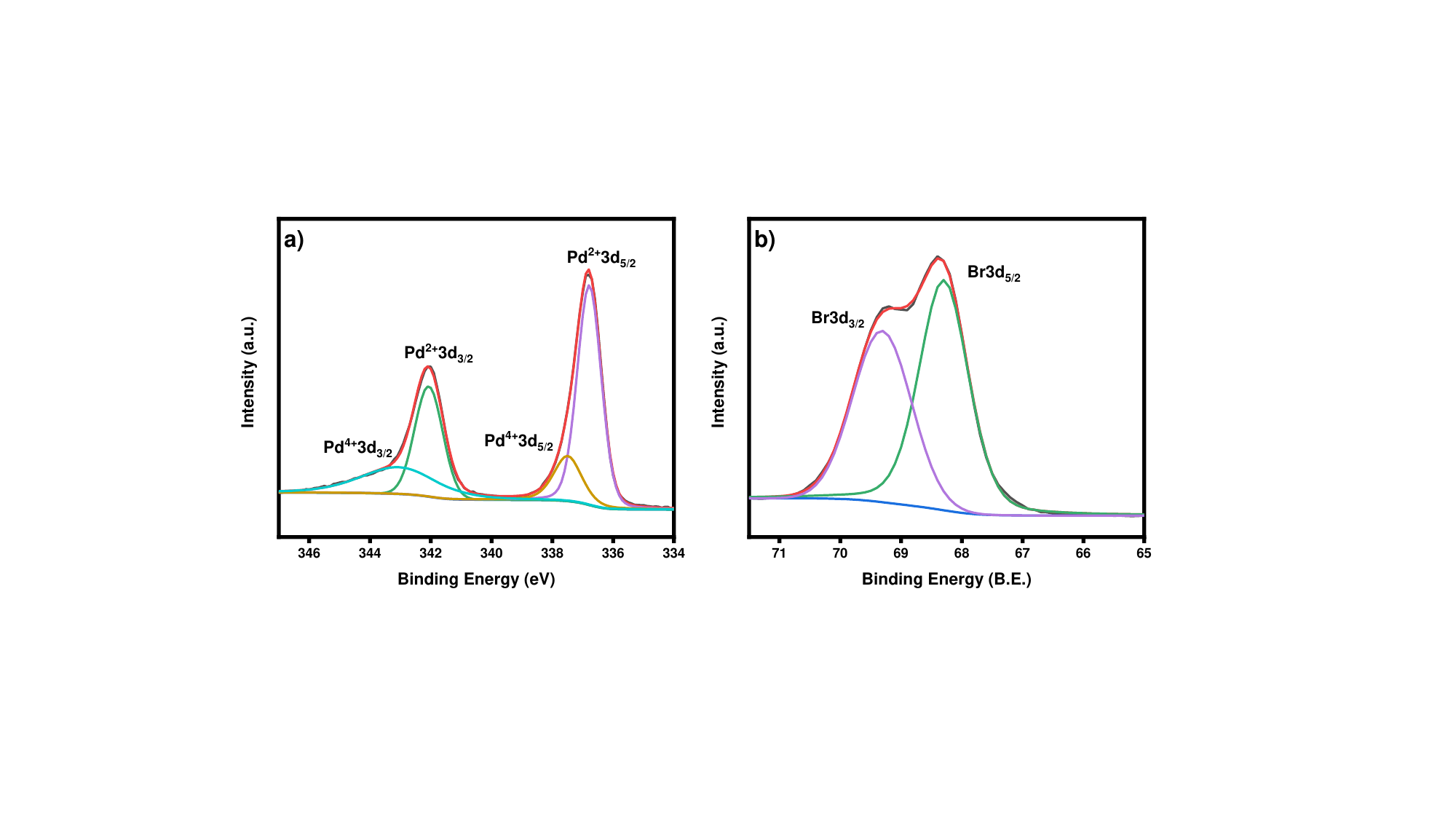}}
\renewcommand{\thefigure}{S\arabic{figure}}
\caption{Core-level 3d spectrum of (a)Pd and (b)Br. Peaks located at 336.8 eV and
342.1 eV correspond to $Pd3d_{5/2}$ and $Pd3d_{3/2}$ spin-orbit components of PdO, respectively. Other two peaks at 337.5 eV and 343.1 eV correspond
to $PdO_2$, indicating the surface oxidation of the material. Peaks located at 68.3 eV and 69.3 eV (Figure (b)) correspond to $Br3d_{5/2}$ and $Br3d_{3/2}$, respectively.
\label{XPS}}
\end{figure} 

\begin{figure}[ht]
\centerline{\includegraphics[scale=0.7, clip]{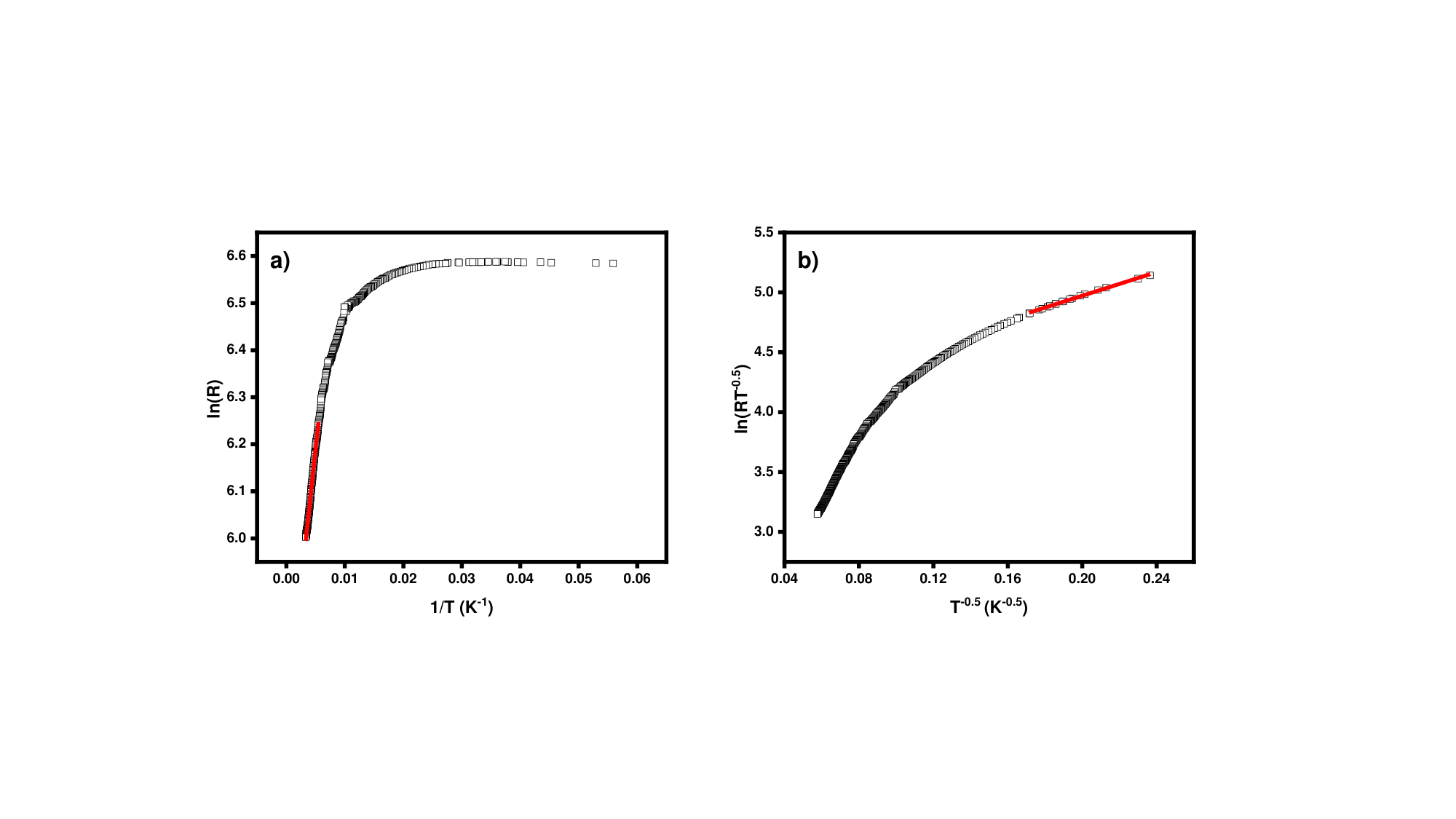}}
\renewcommand{\thefigure}{S\arabic{figure}}
\caption{(a) $ln(R)$ \textit{vs.} $1/T$ graph. At higher temperature range (140-300 K) (the linear fit shown by the red line), the charge transport is mainly dominated by the thermally activated band conduction mechanism. The activation energy ($E_A$) is calculated to be $\approx 10.8 meV$ for a single nanowire of width 55 nm. (b) $ln(RT^{-0.5})$ \textit{vs.} $\frac{1}{\sqrt{T}}$ plot according to VRH model in 1D. The red line shows the linear fit of the plot at low temperature.
\label{Rfitting}}
\end{figure} 

\begin{figure}[ht]
\centerline{\includegraphics[scale=0.7, clip]{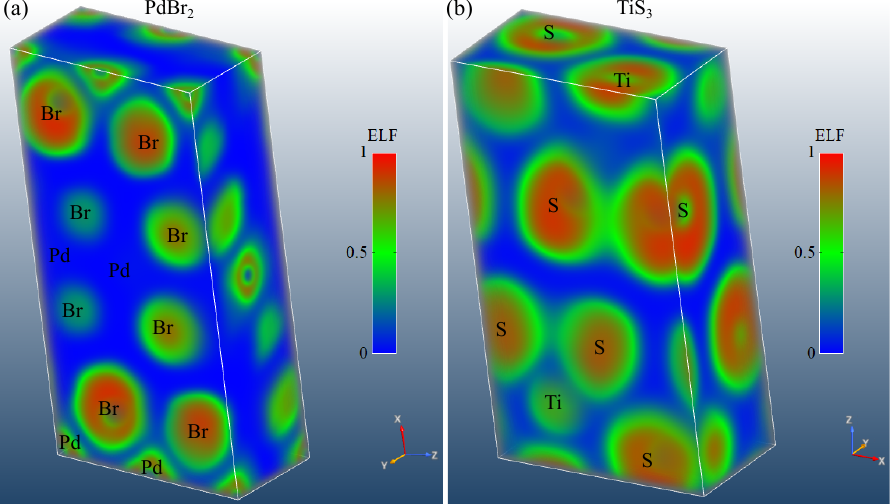}}
\renewcommand{\thefigure}{S\arabic{figure}}
\caption{Calculated electron localization function (ELF) maps for bulk PdBr$_2$ (a) and TiS$_3$ (b). Bulk PdBr$_2$, TiS$_3$ have Monoclinic lattice ($a= 13.04~\AA,~b=3.99~\AA, c=6.62\AA, \alpha= 90^{\circ}, \beta= 102^{\circ}, \gamma= 90^{\circ}$), and ($a= 4.95~\AA,~b=3.38~\AA, c=8.75\AA, \alpha= 90^{\circ}, \beta= 97.61^{\circ}, \gamma= 90^{\circ}$), respectively.
\label{polar}}
\end{figure}